\documentclass[iop]{emulateapj}

\newcommand{\msun} {$M_{\odot}$}

\newcommand{\rsun} {$R_{\odot}$}

\newcommand{\teff} {$T_{\rm eff}$}
\newcommand{\logg}{$\log{g}$}

\newcommand{\omc}{($O-C$)}
\newcommand{\kms}{km s$^{-1}$}
\newcommand{\muhz}{$\mu$Hz}

\usepackage{natbib}
\usepackage{latexsym}
\usepackage{url}
\usepackage[utf8]{inputenc}
\usepackage{fancyref}
\usepackage{hyperref}
\hypersetup{colorlinks,breaklinks,
            urlcolor=[rgb]{0.2,0.2,0.2},  
            linkcolor=[rgb]{0,0,0},  
            citecolor=[rgb]{0,0.4,1.0}}  

\begin{document}

\title{RADIUS CONSTRAINTS FROM HIGH-SPEED PHOTOMETRY OF 20 LOW-MASS WHITE DWARF BINARIES}

\author{J.~J.~Hermes\altaffilmark{1,2}, Warren~R.~Brown\altaffilmark{3}, Mukremin~Kilic\altaffilmark{4}, A.~Gianninas\altaffilmark{4}, Paul~Chote\altaffilmark{5,6}, D.~J.~Sullivan\altaffilmark{5,6}, D.~E.~Winget\altaffilmark{2}, Keaton~J.~Bell\altaffilmark{2}, R.~E.~Falcon\altaffilmark{2}, K.~I.~Winget\altaffilmark{2}, Paul~A.~Mason\altaffilmark{7}, Samuel~T.~Harrold\altaffilmark{2}, and M.~H.~Montgomery\altaffilmark{2}}

\altaffiltext{1}{Department of Physics, University of Warwick, Coventry\,-\,CV4~7AL, United Kingdom}
\altaffiltext{2}{Department of Astronomy, University of Texas at Austin, Austin, TX\,-\,78712, USA}
\altaffiltext{3}{Smithsonian Astrophysical Observatory, 60~Garden~St., Cambridge, MA\,-\,02138, USA}
\altaffiltext{4}{Homer L. Dodge Department of Physics and Astronomy, University of Oklahoma, 440~W.~Brooks~St., Norman, OK\,-\,73019, USA}
\altaffiltext{5}{School of Chemical \& Physical Sciences, Victoria University of Wellington, New Zealand}
\altaffiltext{6}{Visiting Astronomer, Mt. John University Observatory, operated by the University of Canterbury, New Zealand}
\altaffiltext{7}{Department of Physics, University of Texas at El Paso, El Paso, TX\,-\,79968, USA}

\email{j.j.hermes@warwick.ac.uk}

\begin{abstract}

We carry out high-speed photometry on 20 of the shortest-period, detached white dwarf binaries known and discover systems with eclipses, ellipsoidal variations (due to tidal deformations of the visible white dwarf), and Doppler beaming. All of the binaries contain low-mass white dwarfs with orbital periods less than 4\,hr. Our observations identify the first eight tidally distorted white dwarfs, four of which are reported for the first time here, which we use to put empirical constraints on the mass-radius relationship for extremely low-mass ($\leq 0.30$ \msun) white dwarfs. We also detect Doppler beaming in several of these binaries, which confirms the high-amplitude radial-velocity variability. All of these systems are strong sources of gravitational radiation, and long-term monitoring of those that display ellipsoidal variations can be used to detect spin-up of the tidal bulge due to orbital decay.

\end{abstract}

\keywords{binaries: close --- Galaxy: stellar content --- Stars: white dwarfs --- variables: general}

\section{INTRODUCTION}
\label{sec:intro}

White dwarf (WD) stars are galactic fossils that provide a glimpse into the final stages of the evolution of all stars with initial masses below about $7-9$ \msun\ \citep{Dobbie06}. This is the case not only for isolated stars, but also for the large number found in binary systems. WDs thus provide observational boundary conditions on both stellar and binary evolution.

The fate of all binary systems is to eventually be brought together due to the loss of orbital angular momentum carried away by gravitational radiation. This process is usually impractically slow, but the most compact detached binaries ($P_{\rm orb}<6$ hr, $a_{\rm sep} < 1$ \rsun) will merge within a Hubble time. A WD can be uniquely packed into very close orbital configurations with another compact companion and still remain detached: a double-degenerate binary. Such systems will merge to form a variety of exotic objects we see in nature, including AM CVn systems, hydrogen-deficient stars including R Coronae Borealis stars, single subdwarfs, and Supernovae Ia \citep{Iben84,Webbink84,Saio02}.

A growing number of double-degenerate binary systems have been found that contain extremely low-mass (ELM, $M\leq0.30$\msun) WDs, which are by necessity products of binary evolution, since an isolated WD could not have evolved to such a low mass within the age of the Universe. These rare ELM WDs underwent severe mass loss from a close companion, which truncated their evolution and left behind a He-core WD \citep{Iben85}. ELM WDs were first found as companions to millisecond pulsars, but are increasingly found from all-sky surveys like the Sloan Digital Sky Survey (SDSS, e.g., \citealt{Liebert04,BrownELMv}).

He-core WDs provide unique insight into the later stages of close binary evolution and inspiral, and also provide excellent constraints on neutron star masses and equation of state when found as companions to pulsars (e.g., \citealt{vanKerkwijk11,Antoniadis13}). Such work relies on the fundamental parameters for ELM WDs --- especially the mass-radius relationship --- in order to derive cooling ages and to assign masses to these WDs from their spectroscopically determined surface gravities. However, no direct radius measurements for low-mass WDs existed until very recently, and there are still few eclipsing systems known \citep{Steinfadt10a,Parsons11,Vennes11,BrownJ0651,Kilic14}.

\begin{deluxetable*}{lccccccccccc}
\tabletypesize{\scriptsize}
\tablecolumns{12}
\tablewidth{0pc}
\tablecaption{Atmospheric and Binary Parameters of our Low-Mass WDs \label{tab:params}}
\tablehead{
        \colhead{System}&
        \colhead{$T_{\rm eff,1}$}&
        \colhead{$\log{g_1}$}&
        \colhead{{\em $P_{\rm orb}$}}&
        \colhead{$K_1$}&
        \colhead{$M_1$}&
        \colhead{$M_2$}&
        \colhead{$M_2(60\arcdeg)$}&
        \colhead{$\tau_{\rm merge}$}&
        \colhead{$g$}&
        \colhead{$T_{\rm obs}$}&
        \colhead{Ref.}\\
  & (K) & (cm s$^{-2}$) & (days) & (\kms) & (\msun) & (\msun) & (\msun) & (Myr) & (mag) & (hr) & }
\startdata
J065133.34+284423.4   & 16340(260) & 6.81(05) & 0.00886 & 616.9(5.0) & 0.25 &       0.50 & \nodata & $\le 1.2$  & 19.1 & 60.4 & 1,2 \\
J010657.39$-$100003.3 & 16970(260) & 6.10(05) & 0.02715 & 395.2(3.6) & 0.19 & $\ge 0.39$ & 0.51    & $\le 36$   & 19.7 & 14.9 & 3 \\
J163030.58+423305.8   & 16070(250) & 7.07(05) & 0.02766 & 295.9(4.9) & 0.31 & $\ge 0.30$ & 0.38    & $\le 31$   & 19.0 &  7.8 & 4 \\
J105353.89+520031.0   & 16370(240) & 6.54(04) & 0.04256 & 265(15)    & 0.21 & $\ge 0.27$ & 0.34    & $\le 146$  & 19.0 &  4.7 & 5,6 \\
J005648.23$-$061141.6 & 12230(180) & 6.17(04) & 0.04338 & 376.9(2.4) & 0.17 & $\ge 0.46$ & 0.61    & $\le 120$  & 17.4 & 12.2 & 7  \\
J105611.03+653631.5   & 21010(360) & 7.10(05) & 0.04351 & 296.0(7.4) & 0.34 & $\ge 0.40$ & 0.50    & $\le 75$   & 19.8 &  3.0 & 8 \\
J092345.60+302805.0   & 18500(290) & 6.88(05) & 0.04495 & 296.0(3.0) & 0.28 & $\ge 0.37$ & 0.47    & $\le 102$  & 15.7 & 10.8 & 9 \\
J143633.29+501026.8   & 17370(250) & 6.66(04) & 0.04580 & 347.4(8.9) & 0.23 & $\ge 0.46$ & 0.60    & $\le 107$  & 18.2 & 12.5 & 5,6 \\
J082511.90+115236.4   & 27180(400) & 6.60(04) & 0.05819 & 319.4(2.7) & 0.29 & $\ge 0.49$ & 0.64    & $\le 159$  & 18.8 &  7.2 & 8 \\
J174140.49+652638.7   & 10540(170) & 6.00(06) & 0.06111 & 508.0(4.0) & 0.17 & $\ge 1.11$ & 1.57    & $\le 160$  & 18.4 & 13.0 & 10 \\
J075552.40+490627.9   & 13590(280) & 6.13(06) & 0.06302 & 438.0(5.0) & 0.18 & $\ge 0.81$ & 1.13    & $\le 210$  & 20.2 &  5.5 & 9 \\
J233821.51$-$205222.8 & 16620(280) & 6.85(05) & 0.07644 & 133.4(7.5) & 0.26 & $\ge 0.15$ & 0.18    & $\le 970$  & 19.7 &  1.8 & 7  \\
J084910.13+044528.7   & 10290(150) & 6.29(05) & 0.07870 & 366.9(4.7) & 0.18 & $\ge 0.65$ & 0.89    & $\le 440$  & 19.3 & 11.7 & 5 \\
J002207.65$-$101423.5 & 20730(340) & 7.28(05) & 0.07989 & 145.6(5.6) & 0.38 & $\ge 0.21$ & 0.25    & $\le 620$  & 19.8 &  2.2 & 11 \\
J075141.18$-$014120.9 & 15750(250) & 5.49(05) & 0.08001 & 432.6(2.3) & 0.19 & 0.97 & \nodata       & $\le 320$  & 17.5 & 63.2 & 7  \\
J211921.96$-$001825.8 &  9980(150) & 5.71(08) & 0.08677 & 383.0(4.0) & 0.16 & $\ge 0.75$ & 1.04    & $\le 570$  & 20.2 & 11.8 & 2 \\
J123410.36$-$022802.8 & 17800(260) & 6.61(04) & 0.09143 &  94.0(2.3) & 0.23 & $\ge 0.09$ & 0.11    & $\le 2600$ & 17.9 &  8.4 & 11 \\
J074511.56+194926.5   &  8380(130) & 6.21(07) & 0.11165 & 108.7(2.9) & 0.16 & $\ge 0.10$ & 0.12    & $\le 5500$ & 16.5 & 10.9 & 10 \\
J011210.25+183503.7   & 10020(140) & 5.76(05) & 0.14698 & 295.3(2.0) & 0.16 & $\ge 0.62$ & 0.85    & $\le 2700$ & 17.3 & 12.8 & 10 \\
J123316.20+160204.6   & 11700(240) & 5.59(07) & 0.15090 & 336.0(4.0) & 0.17 & $\ge 0.85$ & 1.19    & $\le 2200$ & 19.9 &  5.6 & 9 
\enddata
\tablerefs{ (1) \citet{BrownJ0651}; (2) \citet{HermesJ0651}; (3) \citet{KilicJ0106}; (4) \citet{KilicJ1630}; (5) \citet{Kilic10}; (6) \citet{Mullally09}; (7) \citet{BrownELMv}; (8) \citet{KilicELMiv}; (9) \citet{BrownELMi}; (10) \citet{BrownELMiii}; (11) \citet{KilicELMii}}
\end{deluxetable*}

Searching for more eclipsing systems yields the most precise fundamental parameters for low-mass WDs, but eclipses provide more than simple empirical constraints on WD radii. The times of primary and secondary eclipses can also constrain the orbital eccentricities and mass ratios of the binaries \citep{Kaplan10}, and monitoring the mid-eclipse times provides insight into the orbital evolution of the most compact systems, which are rapidly decaying due to gravitational radiation and tidal effects (e.g., \citealt{Fuller13,Burkart13}). These tidal effects are observationally poorly constrained.

At high enough precision, radius measurements can distinguish different WD hydrogen layer masses (e.g., \citealt{Parsons10}). Measuring the hydrogen layer mass is important for deriving the cooling age for a low-mass WD, since this outermost insulating layer regulates the rate at which the star cools in the stages absent of residual hydrogen burning via the p-p or CNO cycles. Better calibrating cooling ages is essential for using ELM WD companions to millisecond pulsars to constrain the pulsar spin-down age (e.g., \citealt{Kulkarni86}). Improving radius measurements of ELM WDs can also improve distance estimates to systems with pulsar supernova remnants, as the dispersion measure is not always a precise distance indicator \citep{Kaplan13}.

Even coarsely obtaining empirical mass-radius constraints for ELM WDs can help understand the cooling evolution of low-mass WDs. Theoretical models predict that all but the lowest-mass WDs undergo at least one and possibly a series of CNO flashes, as the diffusive hydrogen tail reaches deep enough in the star to ignite CNO burning, which can briefly inflate the star by a factor of hundreds (e.g., \citealt{Panei07,Althaus13}). However, these CNO flashes are so far observationally unconstrained.

Tidally distorted WDs in non-eclipsing systems provide observational constraints on stellar radii, since the amplitude of the photometric variations as a result of tidal distortions scales roughly as $\delta f_{EV} \propto (M_2/M_1)(R_1/a)^3$, where $a$ is the orbital semi-major axis and $R_1$ is the radius of the primary \citep[e.g.,][]{Kopal59}. With spectroscopic constraints on the component masses, ellipsoidal variations can thus yield radii estimates for tidally distorted WDs.

A large number of new ELM WDs have been discovered by the ELM Survey, a targeted spectroscopic search for low-mass WDs from SDSS colors \citep{KilicELMiv,BrownELMv}. The survey has uncovered more than 50 double-degenerate binaries, including four systems with $<$1 hr orbital periods \citep{KilicJ0106,KilicJ1630,BrownJ0651,Kilic20min}. Based on the probability of observing eclipses and other effects that can put fundamental constraints on the physical parameters of these binaries, we have established a follow-up program to photometrically observe the shortest-period systems.

Here we report photometric analysis of the 20 shortest-period binaries from the ELM Survey, all of which have orbital periods $<4$ hr, and most of which fit within orbital separations $<1$ \rsun. These WDs are all in detached, single-lined spectroscopic binaries, suggesting that the unseen companion must be compact; these are all double-degenerate binary systems. In Section 2 we present our observational procedure and describe our methods for characterizing binary variability. Section 3 outlines our results, including our use of the observed ellipsoidal variations to put empirical constraints on the mass-radius relationship for ELM WDs with masses $<0.17$ \msun, and a description of the possibility of using the observed tidal distortions as a probe of orbital decay due to gravitational radiation in these systems. We reserve Section 4 for our conclusions and future work. The Appendix outlines further information about some of our photometrically monitored binaries.

\section{OBSERVATIONS AND METHODS}

\subsection{Target selection}

All 20 of the compact binaries discussed here were discovered through the ELM Survey, whose binary parameters we outline in Table~\ref{tab:params}, in order of increasing orbital period. The effective temperature and surface gravity determinations have been updated to reflect fits to the latest 1D model  atmospheres, as detailed in \citet{Gianninas14b}.

The mass estimates use the most recent evolutionary sequences of He-core WDs of \citet{Althaus13}. We update the merger times to reflect these new primary mass estimates using the formalism of \citet{Landau58}. Since these are all single-lined spectroscopic binaries, $K_1$ refers to the RV semi-amplitude of the visible low-mass WD, and we follow the convention that the primary is the star that is visible, even though the primary is usually the lowest-mass component. In all cases we adopt a systematic uncertainty of $\pm0.02$ \msun\ for $M_1$.

The targets in this sample range in SDSS-$g$ magnitude from $15.7-20.2$ mag. They are predicted to be strong sources of gravitational wave radiation, as all of them will merge within 5.5 Gyr; the 10 shortest-period systems have $P_{\rm orb} < 1.5$ hr and will merge in less than 160 Myr. These binaries all reside within $0.2-3.8$ kpc.

\subsection{Time-series photometry}

The majority of our high-speed photometric observations were obtained at the McDonald Observatory in the three years between 2010~June and 2013~May. For these observations we used the Argos instrument, a frame-transfer CCD mounted at the prime focus of the 2.1 m Otto Struve telescope \citep{Nather04}. Observations were obtained through a {\em BG40} filter to reduce sky noise, which covers a wavelength range of roughly 3000\,\AA\ to 7000\,\AA\ and is centered at 4550\,\AA.

We performed weighted, circular aperture photometry on the calibrated frames using the external IRAF package $\textit{ccd\_hsp}$ \citep{Kanaan02}. We divided the sky-subtracted light curves by the brightest comparison stars in the field to correct for transparency variations, and applied a timing correction to each observation to account for the motion of the Earth around the barycenter of the solar system \citep{Stumpff80,Thompson09}.

Additional observations of J0106$-$1000 were obtained using the GMOS-S instrument mounted on the 8.1 m Gemini-South telescope at Cerro Pach{\'o}n over 4.3 hr in 2011~September and October, under program GS-2011B-Q-52. The first 2.1 hr of observations were obtained through an SDSS-$g$ filter, followed by 2.2 hr of observations through an SDSS-$r$ filter. We performed aperture photometry using DAOPHOT \citep{Stetson87} and used a dozen photometrically constant SDSS point sources in our images for calibration. For these data we used the IDL code of \citet{Eastman10} to apply a barycentric correction.

Extended time-series photometry for J0751$-$0141 was obtained using the Puoko-nui instrument \citep{Chote14} mounted at the Cassegrain focus of the 1.0 m telescope at Mt. John Observatory. These observations were also obtained through a {\em BG40} filter and were reduced using identical procedures as the Argos data from McDonald Observatory.

For two of the faintest targets in our sample, J0755+4906 and J2338$-$2052, we obtained data during a commissioning run using a science camera mounted at the f/5 wavefront sensor of the 6.5 m MMT telescope. This marks some of the first data published taken with this camera. J0755+4906 was also observed using the DIAFI instrument mounted on the 2.7 m Harlan J. Smith telescope at McDonald. Both the MMT and DIAFI data were obtained through an SDSS-$g$ filter, and we performed aperture photometry with $\textit{ccd\_hsp}$. As with J0106$-$1000, we applied a barycentric correction to these observations using the IDL code of \citet{Eastman10}.

For each system, the total duration of our photometric observations ($T_{\rm obs}$) can be found in the second-to-last column of Table~\ref{tab:params}.

\subsection{Characterizing binary variability}
\label{sec:method}

There are six major effects that can cause photometric variability in the primary of a binary system: eclipses, reprocessed light from the secondary (reflection), ellipsoidal variations, Doppler beaming, pulsations, and RV shifts into and out of a narrow-band filter \citep{Robinson87}. We have been fortunate to see all of these effects, save for reflection, in our photometry of ELM WDs.

We proceed to constrain variability using a harmonic analysis, outlined in \citet{HermesJ1741}. In short, we phase our observations to the spectroscopic conjugation and group the data into 100 orbital phase bins. We perform a simultaneous, non-linear least-squares fit with a five-parameter model that includes an offset and defines the amplitude of the (co)sine terms for Doppler beaming ($\sin{\phi}$), ellipsoidal variations ($\cos{2\phi}$), reflection ($\cos{\phi}$), and the first harmonic of the orbital period ($\sin{2\phi}$). The values are detailed in Table~\ref{tab:analysis}, and the stated uncertainties arise from the covariance matrix of this fit.

We generate point-by-point photometric uncertainties using the formalism described in \citet{EH01}, which we adopt in the final binned light curves in Figures~$1-4$. However, this generally underestimates the uncertainties by roughly 40\% --- there are eight systems for which we do not detect any significant variations at the orbital or half-orbital period, and we find these eight systems have, on average, $\chi^2_{\rm red}=\chi^2/\rm{d.o.f.}=1.92$. We rescale the uncertainties for all systems so that $\chi^2_{\rm red}=1.0$, with the expectation that these eight systems only have variability at the orbital period commensurate with the RV-predicted Doppler beaming signal (see Section~\ref{sec:beaming}).

We use the lack of eclipses or a reflection effect to put constraints on the system inclination and maximum temperature of the secondary, respectively, and list these additional constraints in Table~\ref{tab:analysis}. To constrain this inclination angle, we use the He- and CO-core WD mass-radius models of \citet{Althaus13} and \citet{Renedo10} to estimate the radius of the secondary given its minimum mass, which we coarsely arrive at given $i < \sin^{-1}{[(R_1+R_2)/a]}$; the eclipse depth, which roughly scales as $(R_2/R_1)^2$, would be detectable given our observations for all but two systems (J0755+4906 and J1233+1602). We also use this estimate for the secondary radius to constrain the temperature of the secondary using the maximum value of the $\cos{\phi}$ in Table~\ref{tab:analysis} and the expected amplitude of a reflection effect approximated by
$\delta f = \frac{17}{16}(R_1/a)^2[\frac{1}{3} + \frac{1}{4}(R_1/a)](T_2/T_1)^4(R_2/R_1)^4$
\citep{Kopal59}.

\begin{deluxetable*}{lcccccccccccc}
\tabletypesize{\scriptsize}
\tablecolumns{13}
\tablewidth{0pc}
\tablecaption{Light Curve Analysis of 20 Merging Low-Mass WD Binaries \label{tab:analysis}}
\tablehead{
        \colhead{Object}&  
        \colhead{$P_{\rm orb,fold}$}&  
        \colhead{$DB_{\rm exp}$}& 
        \colhead{$\sin(\phi)$}&    
        \colhead{$\cos(2\phi)$}&   
        \colhead{$\cos(\phi)$}&    
        \colhead{$\sin(2\phi)$}&   
        \colhead{$\chi^2_{\rm red}$}&
        \colhead{$T_2$}&
        \colhead{$i$}&
        \colhead{$M_2$}&
        \colhead{$u_1$}&
        \colhead{$\tau_1$}\\
  & (min) & (\%) & (\%) & (\%) & (\%) & (\%) &  & (kK) & (deg) & (\msun) }
\startdata
J0651+2844   & 12.7534424 & 0.47 & 0.56(07) & 4.01(07) & 0.02(07) & 0.01(07) & \nodata & $8.7(0.5)$ & $84.4(2.3)$ & $0.50$ & 0.39 & 0.566\\
J0106$-$1000 & 39.10406   & 0.29 & 0.23(12) & 1.76(12) & 0.19(12) & 0.02(12) & 1.00 & $<21$ & $<76.6$ & $>0.41$ & 0.40 & 0.552 \\
J1630+4233   & 39.830     & 0.22 & 0.32(09) & 0.02(09) & 0.00(09) & 0.02(09) & 0.91 & $<17$ & $<82.8$ & $>0.66$ & \nodata & \nodata \\
J1053+5200   & 61.286     & 0.20 & 0.22(13) & 0.35(13) & 0.02(13) & 0.15(13) & 0.92 & $<21$ & $<82.0$ & $>0.27$ & \nodata & \nodata \\
J0056$-$0611 & 62.466700  & 0.35 & 0.04(06) & 0.53(06) & 0.01(06) & 0.05(06) & 1.02 & $<13$ & $<81.8$ & $>0.47$ & 0.46 & 0.700 \\
J1056+6536   & 62.654     & 0.17 & 0.68(29) & 0.34(29) & 0.00(29) & 0.19(28) & 0.95 & $<36$ & $<84.9$ & $>0.34$ & \nodata & \nodata \\
J0923+3028   & 64.8482    & 0.21 & 0.43(03) & 0.08(02) & 0.07(02) & 0.16(02) & 0.64 & $<25$ & $<84.6$ & $>0.37$ & \nodata & \nodata \\
J1436+5010   & 65.95      & 0.25 & 0.35(05) & 0.12(05) & 0.01(05) & 0.07(05) & 0.67 & $<23$ & $<84.4$ & $>0.46$ & \nodata & \nodata \\
J0825+1152   & 83.7936    & 0.18 & 0.43(14) & 0.04(14) & 0.04(14) & 0.19(14) & 0.92 & $<52$ & $<84.8$ & $>0.50$ & \nodata & \nodata \\
J1741+6526   & 87.9984    & 0.54 & 0.50(07) & 1.30(07) & 0.11(07) & 0.00(07) & 1.01 & $<31$ & $<84.4$ & $>1.12$ & 0.54 & 0.791 \\
J0755+4906   & 90.749     & 0.38 & 0.36(29) & 1.05(29) & 0.82(30) & 0.22(30) & 0.84 & $<58$ & $<90.0$ & $>0.81$ & \nodata & \nodata \\
J2338$-$2052 & 110.07     & 0.10 & 0.00(16) & 0.30(15) & 0.19(16) & 0.01(16) & 0.98 & \nodata & $<83.8$ & $>0.15$ & \nodata & \nodata \\
J0849+0445   & 113.2013   & 0.40 & 0.78(13) & 0.41(13) & 0.00(13) & 0.03(12) & 0.66 & $<24$ & $<85.7$ & $>0.66$ & \nodata & \nodata \\
J0022$-$1014 & 115.04     & 0.10 & 0.05(35) & 0.01(35) & 0.00(37) & 0.12(36) & 1.03 & \nodata & $<86.0$ & $>0.21$ & \nodata & \nodata \\
J0751$-$0141 & 115.21814  & 0.33 & 0.25(03) & 3.20(03) & 0.20(03) & 0.00(03) & \nodata & $<45$ & $85.4(9.4)$ & $0.97$ & 0.41 & 0.581 \\
J2119$-$0018 & 124.949    & 0.42 & 0.71(15) & 1.44(15) & 0.14(15) & 0.08(15) & 1.02 & $<27$ & $<82.8$ & $>0.76$ & 0.52 & 0.265 \\
J1234$-$0228 & 131.66     & 0.07 & 0.07(06) & 0.13(06) & 0.09(06) & 0.12(06) & 0.91 & \nodata & $<71.5$ & $>0.10$ & \nodata & \nodata \\
J0745+1949   & 161.9298   & 0.15 & 0.00(04) & 1.49(04) & 0.07(04) & 0.01(04) & 0.98 & \nodata & $<72.5$ & $>0.11$ & 0.58 & 0.310 \\
J0112+1835   & 211.55545  & 0.33 & 0.00(03) & 0.32(03) & 0.04(03) & 0.00(03) & 1.02 & $<24$ & $<85.3$ & $>0.63$ & 0.51 & 0.826 \\
J1233+1602   & 217.30     & 0.33 & 0.07(21) & 0.61(22) & 0.22(22) & 0.19(22) & 0.90 & $<57$ & $<90.0$ & $>0.85$ & \nodata & \nodata 
\enddata
\end{deluxetable*}

Additionally, we have computed Fourier transforms (FTs) of our time-series photometry, which allows us to search for any variability, such as pulsations, not at a harmonic of the orbital period. In some cases, this Fourier analysis has allowed us to refine the orbital period of the system from the less-sampled RV observations. Usually, though, the periods are so long and the data coverage so sparse that there is too much alias structure around the peaks of interest to justifiably refine the orbital period. Since we have signals that may occur at both $\cos{\phi}$ and $\sin{\phi}$, our Monte Carlo analysis yields a more reliable estimate for the amplitude of reflection and Doppler beaming, respectively. Comparing these FTs to the FT of the brightest comparison star in the field allows us to check for any coincident peaks that may be the result of atmospheric variability or instrumental effects. We have also computed significance levels based on $\langle {\rm A}\rangle$, the average amplitude of the FT within a 1000 \muhz\ region centered at 2000 \muhz.

\section{RESULTS AND BINARY PHYSICAL PARAMETERS}

Our photometric observations constrain the 20 shortest-period ELM WD binaries, all with orbital periods $<4$ hr. We visually represent our results, in order of increasing orbital period, in Figures~$1-4$. We include the orbital periods we have used to fold the light curves in Table~\ref{tab:analysis}.

\begin{figure*}
\centering{\includegraphics[width=\textwidth]{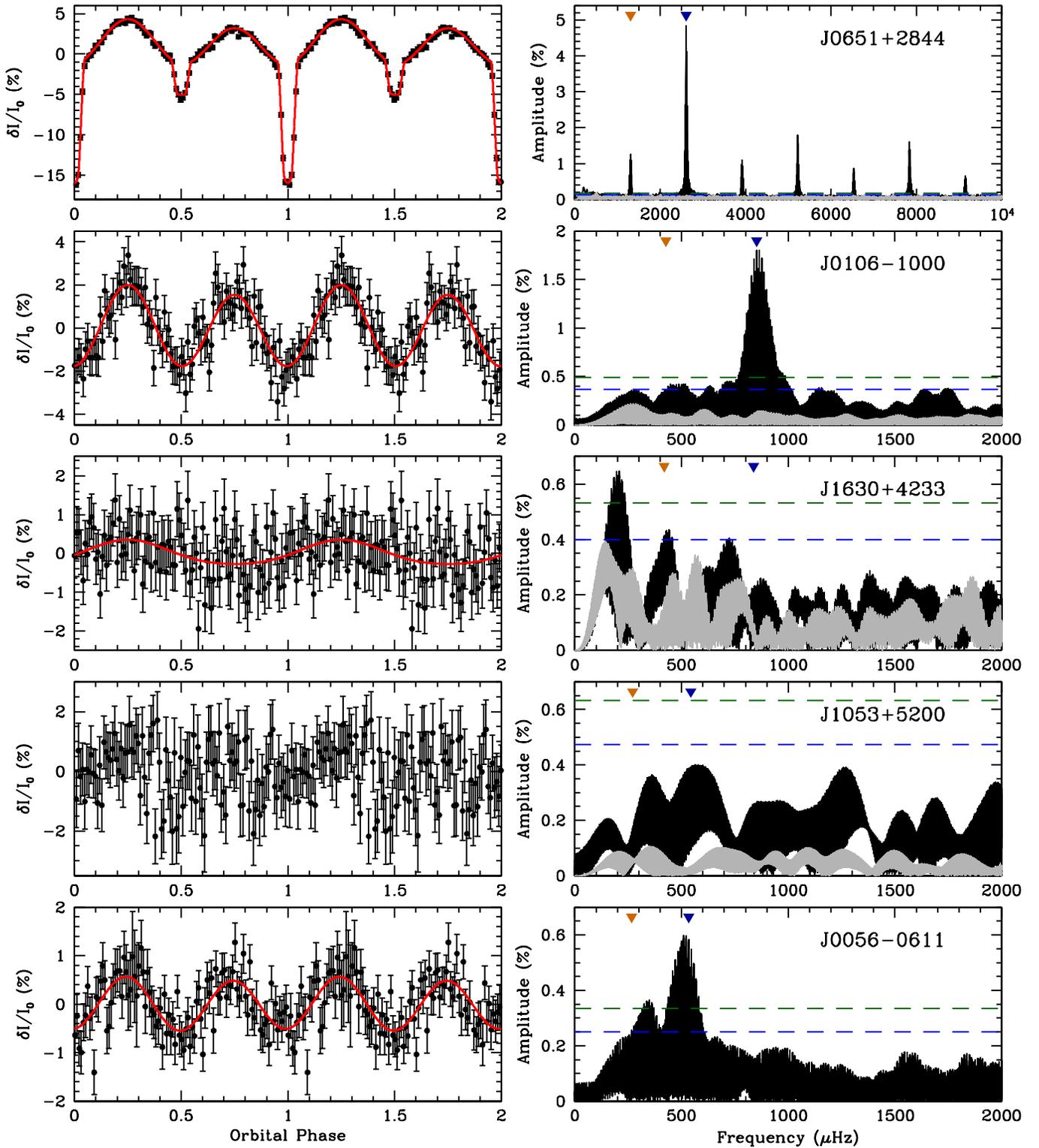}}
\caption{High-speed photometry of five ELM WDs in compact binaries. The left panels show the optical light curves, binned into 100 phase points, folded at the orbital period, and repeated for clarity. The solid red line displays our best-fit model, where appropriate. The right panels show an FT of the target (black) and brightest comparison star (grey). The orange and blue triangles show the orbital period and half-orbital period, respectively. These binaries have orbital periods of 12.8 min (J0651+2844), 39.1 min (J0106$-$1000), 39.8 min (J1630+4233), 61.3 min (J1053+5200), and 62.5 min (J0056$-$0611). The dashed green and blue lines show the 4$\langle {\rm A}\rangle$ and 3$\langle {\rm A}\rangle$ significance level in the FT, respectively. There was only one comparison star in the field for J0056$-$0611. \label{fig:binary1}}
\end{figure*}

\begin{figure*}
\centering{\includegraphics[width=\textwidth]{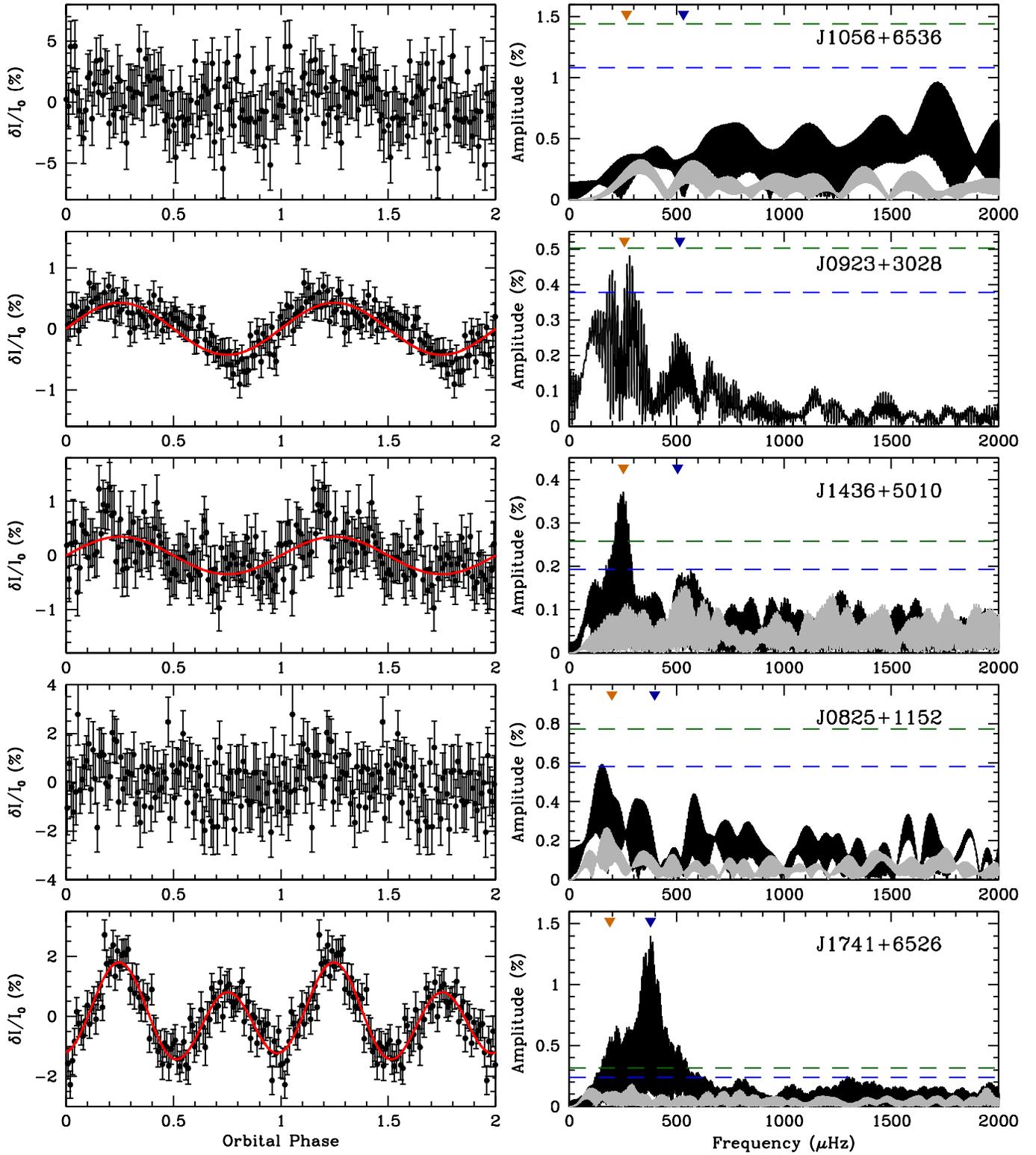}}
\caption{The same as Figure~\ref{fig:binary1} but for five additional compact WD binaries. These systems have orbital periods of 62.5 min (J1056+6536), 64.7 min (J0923+3028), 66.0 min (J1436+5010), 83.8 min (J0825+1152), and 88.0 min (J1741+6526). There was only one comparison star in the field for J0923+3028. \label{fig:binary2}}
\end{figure*}

\begin{figure*}
\centering{\includegraphics[width=\textwidth]{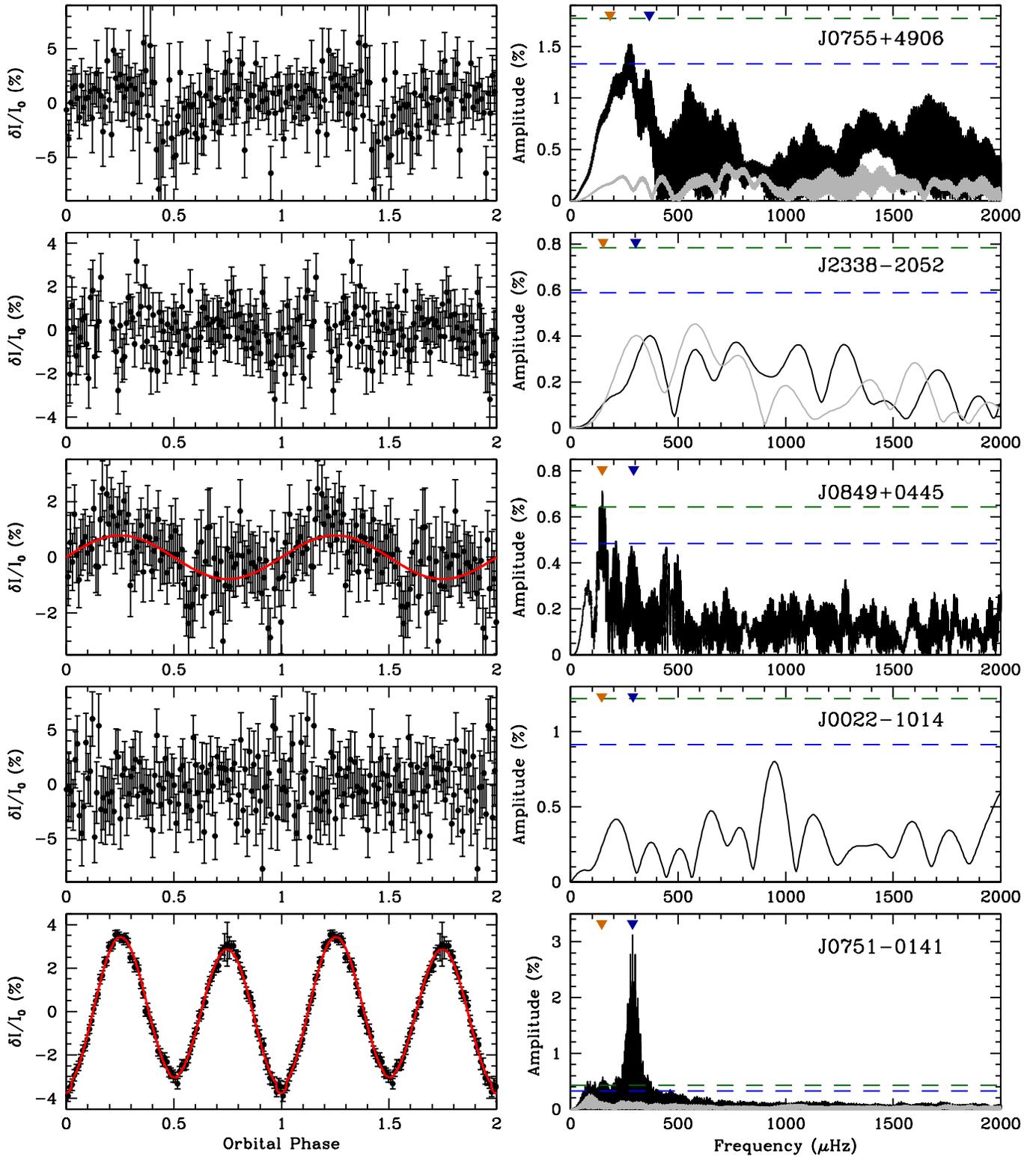}}
\caption{The same as Figure~\ref{fig:binary1} but for five additional compact WD binaries. These systems have orbital periods of 90.7 min (J0755+4906), 110.1 min (J2338$-$2052), 113.3 min (J0849+0445), 115.0 min (J0022$-$1014), and 117.2 min (WD0751$-$0141). There was only one comparison star in the field for J0849+0445 and J0022$-$1014. \label{fig:binary3}}
\end{figure*}

\begin{figure*}
\centering{\includegraphics[width=\textwidth]{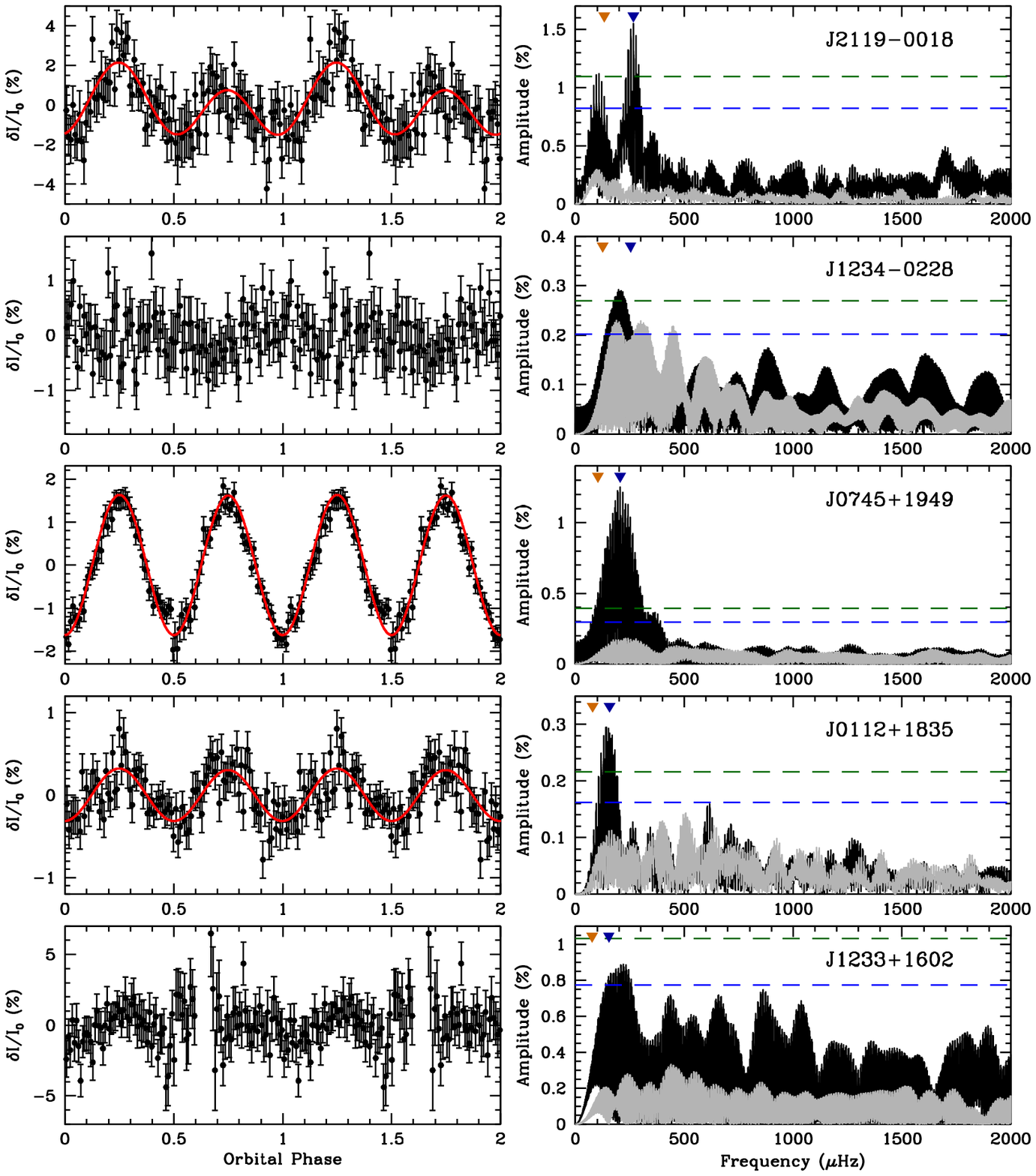}}
\caption{The same as Figure~\ref{fig:binary1} but for five additional compact WD binaries. These systems have orbital periods of 124.9 min (J2119$-$0018), 131.7 min (J1234$-$0228), 161.9 min (J0745+1949), 211.7 min (J0112+1835), and 217.3 min (J1233+1602). \label{fig:binary4}}
\end{figure*}

J0651+2844 is the most compact detached binary known, and has been studied extensively since its discovery in 2011~March; we have detected the signature of orbital decay due to gravitational radiation by monitoring the rapid change in mid-eclipse times \citep{HermesJ0651}. We display here in Figure~\ref{fig:binary1} only our data from 2012~January.

This unique system evidences the three main ways in which compact WD binaries can exhibit photometric variability: deep primary eclipses at $\phi=0$, secondary eclipses at $\phi=0.5$, ellipsoidal variations from tidal distortion of the primary WD peaking twice each orbit, and Doppler beaming at the orbital period, manifest as the higher asymmetry in the ellipsoidal variations at $\phi=0.25$. The red curve in Figure~\ref{fig:binary1} is the best model fit to the data, described in \citet{HermesJ0651}.

The Fourier transform of this data orients us for the other systems. The highest peak in the FT occurs at the half-orbital period, denoted by the dark blue inverted triangle, which serves to reproduce the high-amplitude ellipsoidal variations. There is also a significant peak at the orbital period, denoted by the orange inverted triangle, primarily corresponding to the Doppler beaming signal. Finally, the comb of peaks at harmonics of the orbital period are a Fourier series reproducing the deep eclipses.

We include in Table~\ref{tab:analysis} the full results from our harmonic analysis of the high-speed photometry of our 20 low-mass WD binaries. This harmonic analysis yields amplitudes for Doppler beaming ($\sin \phi$), the dominant component of ellipsoidal variations ($\cos 2\phi$), reflection ($\cos \phi$), and the first harmonic of the orbital period ($\sin 2\phi$). We can use our photometric observations to provide independent constraints on the low-mass WD mass-radius relationship, which we update in Section~\ref{sec:massrad}. Additionally, we can monitor the orbital evolution using these ellipsoidal variations, as we discuss in Section~\ref{sec:gravwaves}.

\subsection{Observed Doppler beaming signals}
\label{sec:beaming}

Doppler beaming introduces a detectable modulation in the flux of a binary star that is approaching or receding, with an expected amplitude primarily dictated by the radial velocity of the source \citep{Zucker07}. The dominant component of the effect is due to aberration of the high-velocity source, although there is also a dependence on the frequency of the emitted radiation with the source velocity \citep{Bloemen11}. 

Observing Doppler beaming in binaries is still new, but by no means novel. The first ground-based detection involved the WD+WD binary NLTT 11748 \citep{Shporer10}, although there was marginal evidence for the effect in the short-period sdB+WD binary KPD 1930+2752 \citep{Maxted00}. The effect is now routinely observed using high-quality, space-based photometry (e.g., \citealt{vankerkwijk10}).

Nine of the 20 systems in our sample display a significant modulation at the orbital period commensurate with Doppler beaming. Since we already know the radial-velocity amplitude of these systems from spectroscopy, detecting this signal yields little new information about our binaries. However, it does help calibrate our photometric uncertainties. We can generally predict the expected Doppler beaming signal to $\pm0.02$\% relative amplitude, which we list in Table~\ref{tab:analysis}, and we compare this to the results of our harmonic analysis. For all but three systems, the observed $\sin{\phi}$ term matches within 3$\sigma$ the expected Doppler beaming amplitude, which is given in Table~\ref{tab:analysis}. We use these Doppler beaming signal predictions to rescale the photometric uncertainties such that $\chi^2_{\rm red}=1.0$.

However, this is not an entirely robust approach, since there is occasionally long-timescale transparency variations or other atmospheric variability contaminating some of the expected signals, which sometimes overlap with the orbital period and thus the Doppler beaming signal. For example, one anomalous system (J0923+3028) has an observed $\sin{\phi}$ term more than twice the expected value, but the photometry may be influenced by long-period atmospheric variability; there is a comparable signal at 1.4 hr which is longer than the 1.08-hr orbital period.

\subsection{Inferred radii from ellipsoidal variations}
\label{sec:radius}

In the cases for which we see tidal distortions, the dominant modulation occurs when the larger face comes into view twice per rotation, effectively a $\cos{2\phi}$ modulation of the rotation rate of the WD tidal bulge, which would be the orbital period for a synchronized system ($\phi=0-2\pi$ represents one full orbit). Since tidal distortions do not cause a perfectly ellipsoidal shape, we treat the ellipsoidal variations as harmonics to the first four $\cos{\phi}$ terms, as derived in \citet{Morris93}. They showed that the ellipsoidal variation amplitude is dominated by
$$
L(\phi) / L_0 = \frac{-3 \, (15 + u_1) \, (1 + \tau_1) \, q \, (R_1 / a)^3 \, \sin^2 i}{20 \, (3-u_1)} \cos(2\phi)
$$
where all terms can be written in terms of the inclination ($i$) given that we can determine the mass ratio ($q=M_2/M_1$) and semimajor axis of the system ($a$) from the spectroscopy and that we can assume reasonable values for the linear limb-darkening ($u_1$) and gravity-darkening ($\tau_1$) coefficients for the primary. We note that this formalism is only valid in the regime when the tidally distorted star is rotating at or near the orbital period, as shown by \citet{Bloemen12}. For now, we assume this is valid. 

We adopt limb-darkening coefficients calculated by \citet{Gianninas13}, which we list in Table~\ref{tab:analysis}. For all WDs with \teff\ $> 10{,}000$ K, we assume the surface is purely radiative and calculate the gravity-darkening coefficients using the formalism outlined in \citet{Morris85}, where $\beta=0.25$. This assumption is likely valid given our theoretical (and growingly empirical) blue edge for pulsating ELM WDs found in Figure~5 of \citet{HermesELMV23}; these pulsations are driven by a hydrogen partial-ionization zone, which coincides with the onset of a deepening surface convection zone. Our adopted gravity-darkening coefficients are included in Table~\ref{tab:analysis}.

Eight systems in our sample show a significant $\cos(2\phi)$ variation in the light curve. For each, we have also calculated the amplitudes of the third- and fourth-cosine harmonics, and find they are insignificant, within the uncertainties, as we would expect from the predicted amplitude ratios of \citet{Morris93}. For all systems, we do not observe these higher-order harmonics to have amplitudes 2$\sigma$ above the least-squares uncertainties; in fact, we do not expect these harmonics to have amplitudes above 0.13\% relative amplitude for any of our systems.

We can rewrite the equation of \citet{Morris93} characterizing the amplitude of the ellipsoidal variations ($A_{\rm EV}$) by recasting the semimajor axis using Kepler's third law:

\begin{equation}
A_{\rm EV} = \frac{3 \pi^2 (15 + u_1)(1 + \tau_1) M_2 R_1^3 \sin^2{i}}{5 P_{\rm orb}^2 (3 - u_1) G M_1 (M_1+M_2)}
\end{equation}

Additionally, we have spectroscopic constraints on the system. Dynamically, we know the mass function from previous time-series spectroscopic observations:

\begin{equation}
f_1(M_2) = \frac{P_{\rm orb} K_1^3}{2 \pi G} = \frac{M_2^3 \sin^3{i}}{(M_1+M_2)^2}
\end{equation}

We also know the primary surface gravity ($g_1$) from atmosphere models fit to its summed spectrum:

\begin{equation}
g_1 = \frac{G M_1}{R_1^2}
\end{equation}

Thus we have three equations and four unknowns ($M_1,M_2,R_1,i$). To draw out the line of mass-radius constraints for each ELM WD, we perform $10{,}000$ Monte Carlo simulations. We draw a random inclination using the proper distribution of random orientations, as well as a random value from within the measured probability distribution for $P_{\rm orb}, K_1, \log{g_1}, A_{\rm EV}$, and solve the system of three equations. We reject any solutions which have $M_1>1.4$ \msun\ or $M_2>3.0$ \msun, but do not impose any inclination constraints.

\subsection{Constraining the low-mass WD mass-radius relation with ellipsoidal variations}
\label{sec:massrad}

\begin{figure*}
\centering
\includegraphics[width=\textwidth]{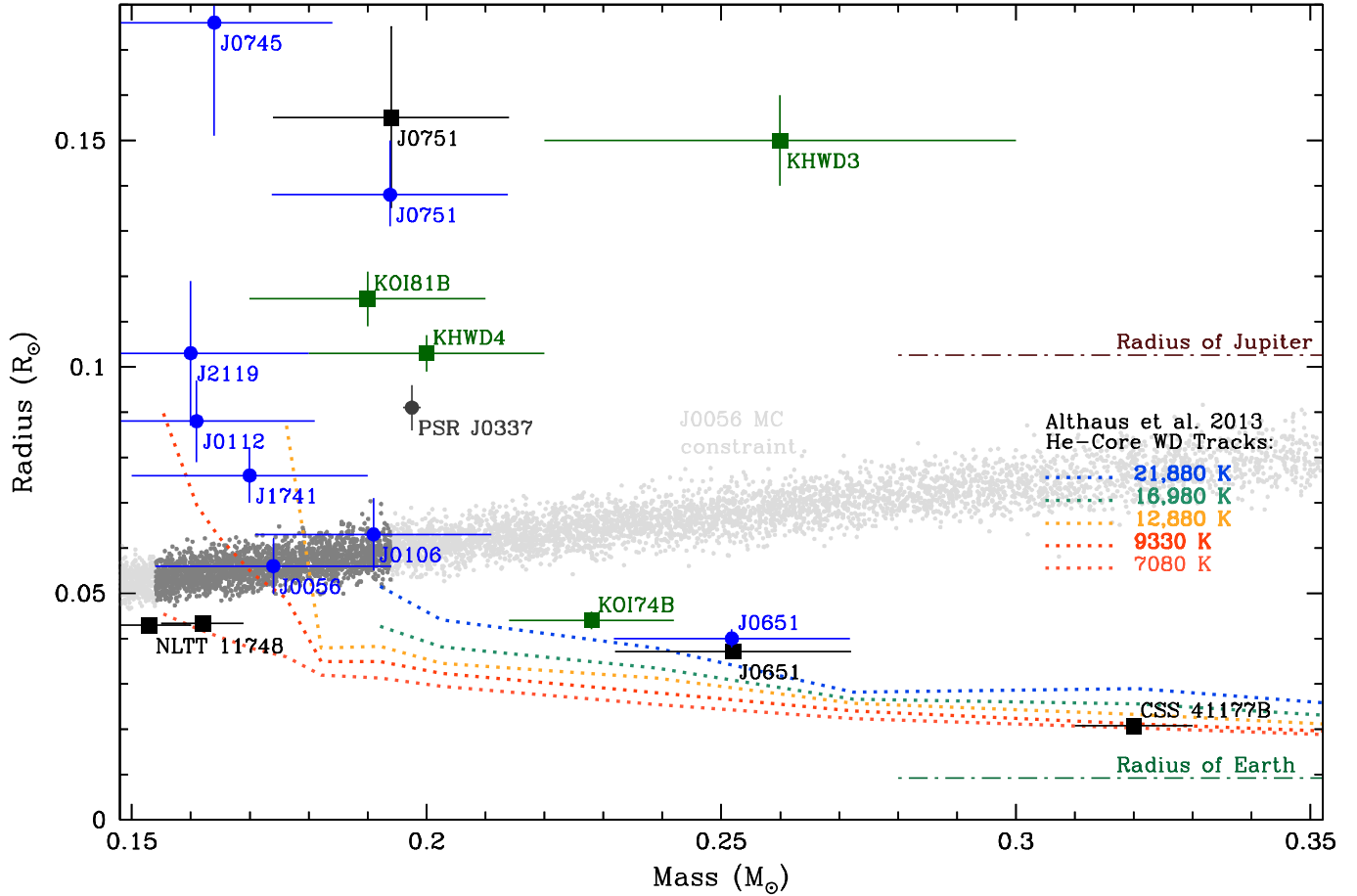}
\caption{Observed mass-radius constraints for low-mass (He-core) WDs. The blue points mark the results of our analysis using the ellipsoidal variations of eight tidally distorted WDs. For one star (J0056$-$0061) we show in gray the full results of our Monte Carlo simulation, and highlight in darker gray the results within the spectroscopically determined mass. Black squares represent WD+WD eclipsing systems described in the text. The overlapping parameters derived from light curve fits to the eclipses in the tidally distorted systems J0651+2844 and J0751$-$0141 verify our method. We also include eclipsing low-mass WD systems which may be bloated because their companions are A stars as dark green squares, and the well-constrained WD companion to PSR J0337+1715 as dark gray point \citep{Kaplan14b}. To guide the eye we include the terminal cooling tracks for theoretical models for He-core WDs from \citet{Althaus13}, which cover a range of temperatures. \label{fig:massradius}} 
\end{figure*}

Our Monte Carlo simulations draw out a series of allowed values along the $M_1-R_1$ plane given the observed ellipsoidal variations. As an example, we display in Figure~\ref{fig:massradius} the full output for J0056$-$0611 as light grey points. To further constrain the radius estimate, we highlight only the points within 0.02 \msun\ (our adopted systematic uncertainty) of the mass adopted by matching the \teff\ and \logg\ to the models of \citet{Althaus13}. The distribution of radii from our Monte Carlo simulations are symmetric for each star constrained within this mass range, so the adopted radius estimates listed in Table~\ref{tab:EVconstraints} are found from a Gaussian fit to this distribution of radii.

The most precise constraints on the radii of low-mass WDs come from detached eclipsing systems. Fortunately, there are now six known low-mass WDs in eclipsing binaries: NLTT 11748 \citep{Steinfadt10a,Kilic10,Kaplan14}, CSS 41177 A\&B \citep{Parsons11,Bours14}, GALEX J1717+6757 \citep{Vennes11}, J0651+2844 \citep{BrownJ0651,HermesJ0651}, and J0751$-$0141 \citep{Kilic14}. We include CSS 41177B \citep{Bours14} and the two higher-mass solutions for NLTT 11748 \citep{Kaplan14} as black squares in Figure~\ref{fig:massradius}; the other systems either do not have stringent enough constraints on the WD radius or have a mass too high to display in this figure.

Two of the systems exhibiting ellipsoidal variations (J0651+2844 and J0751$-$0141) are also eclipsing, and we include in Figure~\ref{fig:massradius} black squares corresponding to the radius values derived from light curve models to the eclipses. There is excellent agreement between the results of our Monte Carlo simulations and the light curve fits, which helps validate our method. 

There are four additional low-mass WDs in eclipsing binaries with A-star companions, all discovered in the {\em Kepler} field. These WDs are KOI81B \citep{vankerkwijk10,Rowe10}, KOI74B \citep{vankerkwijk10,Bloemen12}, KHWD3 \citep{Carter11}, and KHWD4 \citep{Breton11}. We mark these low-mass WDs as green squares in Figure~\ref{fig:massradius} to differentiate between the other eclipsing WDs because the A-star companions could contribute to inflating the radius of the WD \citep{Carter11}.

We also include in Figure~\ref{fig:massradius} the theoretical mass-radius relations of He-core WDs from \citet{Althaus13}. These tracks generally show that the WD radius increases with increasing \teff\ and decreasing mass. Of note, such low-mass WDs ($<0.18$ \msun) are expected to quiescently burn hydrogen and are not theoretically predicted to undergo CNO flashes; for example, the large jump in radius for the low-mass end of the $12{,}880$~K isotherm in Figure~\ref{fig:binary1} demonstrates the expectedly larger radius for a 0.1762 \msun\ model, which does not undergo CNO flashes and thus has a more massive residual hydrogen layer.

Our tidally distorted ELM WDs are among the lowest-mass WDs with radius constraints, so our observational results fill an important and untested region of the mass-radius relation. Six of our eight radius measurements are generally consistent with the models of \citet{Althaus13}. In some cases (notably J0106$-$1000 and J0651+2844), the observed radii are slightly larger than expected given their spectroscopic mass and temperature.

However, two outliers have radii significantly larger than expected from the He-core WD models: J0751$-$0141 and J0745+1949. It is possible that these two WDs are not on their final cooling track, but are instead in another part of their evolution, perhaps recently undergoing a CNO flash. Assuming the surface of J0745+1949 is radiative and adopting a larger gravity-darkening coefficient for this 8380 K WD cannot explain this discrepancy; adopting $\tau_1 = 0.967$ still yields $R_1 = 0.153^{+0.055}_{-0.020}$ \rsun.

It is notable that J0745+1949 is one of the most metal-rich WDs known \citep{Gianninas13b}, which could possibly be the result of mixing induced by a recent CNO flash. If so, the mass determined from the \teff\ and \logg\ may not accurately represent the WD mass, which was adopted assuming the WD was on its terminal cooling track. Higher-mass models indeed cross the same position in \teff\ and \logg\ space while undergoing CNO flashes before their terminal cooling track. There is mounting evidence that many low-mass WDs do not appear to be on a terminal cooling track, especially those that are companions to millisecond pulsars (e.g., \citealt{Kaplan13,Kaplan14b}). There are likely still unexplained complexities to the evolution of low-mass WDs.

If we do not restrict our Monte Carlo simulation analysis of J0745+1949 by the primary mass of 0.164 \msun, we instead find $M_1 = 0.38^{+0.28}_{-0.22}$ \msun, $R_1 = 0.25^{+0.10}_{-0.07}$ \rsun, and $M_2 = 0.19^{+0.16}_{-0.09}$ \msun. However, this would suggest our \logg\ estimate from spectroscopy is off by more than 0.8 dex, which is highly unlikely. The radius of a 0.363 \msun\ model WD in the throes of a CNO flash can change by more than 0.15 \rsun\ in less than a year \citep{Althaus13}, which would cause a clear change in the amplitude of the ellipsoidal variations, so follow-up photometry of J0745+1949 could constrain this scenario.

\begin{deluxetable*}{lccccccc}
\tabletypesize{\scriptsize}
\tablecolumns{8}
\tablewidth{0pc}
\tablecaption{Parameters Constrained from Monte Carlo Simulations Using the Observed Ellipsoidal Variations \label{tab:EVconstraints}}
\tablehead{
        \colhead{Object}&   
        \colhead{$M_1$}& 
        \colhead{$R_1$}&   
        \colhead{$M_2$}& 
        \colhead{$i$}&   
        \colhead{$dP_{\rm EV}/dt_{\rm GR}$}&  
        \colhead{$\tau_{\rm detect}$} &
        \colhead{$T_{\rm 0, ELV}$} \\
  &  (\msun) &  (\rsun) & (\msun) & (deg) & ($10^{-13}$ s s$^{-1}$) & (yr) & (${\rm BJD_{TDB}}$) }
\startdata
J0651+2844   & 0.252 & $0.040\pm0.002$ & $0.50^{+0.04}_{-0.01}$ & $82.7^{+7.3}_{-8.4}$ & $-39.8^{+2.5}_{-0.7}$  & $<1$     & 2455955.1734648(37) \\
J0106$-$1000 & 0.191 & $0.063\pm0.008$ & $0.50^{+0.43}_{-0.11}$ & $60.3^{+28.7}_{-19.5}$ & $-9.6^{+5.6}_{-2.0}$  & $11$ & 2455533.57568(11) \\
J0056$-$0611 & 0.174 & $0.056\pm0.006$ & $0.80^{+0.63}_{-0.30}$ & $49.7^{+22.3}_{-12.8}$ & $-2.9^{+14.0}_{-0.4}$  & $37$ & 2455891.62845(42) \\
J1741+6526   & 0.170 & $0.076\pm0.006$ & $1.16^{+0.41}_{-0.05}$ & $78.3^{+11.7}_{-15.8}$ & $-2.1^{+0.3}_{-0.1}$  & $41$ & 2455686.79210(21) \\
J0751$-$0141 & 0.194 & $0.138^{+0.012}_{-0.007}$ & $1.02^{+0.38}_{-0.05}$ & $77.3^{+12.7}_{-17.2}$ & $-1.3^{+0.4}_{-0.1}$  & $34$ & 2455960.660518(54) \\ 
J2119$-$0018 & 0.160 & $0.103\pm0.016$ & $0.80^{+0.44}_{-0.10}$ & $75.1^{+14.9}_{-20.6}$ & $-0.8^{+0.3}_{-0.1}$  & $150$ & 2455769.84065(66) \\
J0745+1949   & 0.164 & $0.176^{+0.090}_{-0.025}$ & $0.14^{+0.13}_{-0.07}$ & $63.2^{+26.8}_{-32.4}$ & $-0.14^{+0.10}_{-0.06}$ & $180$ & 2456245.94304(51) \\
J0112+1835   & 0.161 & $0.088\pm0.009$ & $0.70^{+0.45}_{-0.11}$ & $70.3^{+19.7}_{-19.2}$ & $-0.32^{+0.14}_{-0.04}$ & $470$ & 2455808.79023(89)
\enddata
\end{deluxetable*}

\subsection{Monitoring for the effects of gravitational radiation}
\label{sec:gravwaves}

The ellipsoidal variations in the shortest-period systems in our sample provide a unique opportunity to act as a stable clock which can be used to monitor any changes to the system as a result of orbital decay from the emission of gravitational wave radiation. In each case, the ellipsoidal variations show that the tidal bulge of the primary is synchronized with the orbital period, to the limit of our uncertainties. As that orbital period shrinks with the emission of gravitational waves, this tidal bulge will spin up, and the period of the ellipsoidal variations will decrease, which we can detect by monitoring the arrival times of the ellipsoidal variations.

Some of these systems are so compact that it is possible to detect the influence of gravitational waves within a decade or less. We have already established that such a monitoring campaign is possible: We have used the time-of-minima of the ellipsoidal variations in the 12.75-min binary J0651+2844 as an independent clock with which to detect the rapid orbital decay due to gravitational wave radiation \citep{HermesJ0651}.

Our Monte Carlo simulations provide additional constraints on the most likely distribution of system inclinations and companion masses, which we include in Table~\ref{tab:EVconstraints}. These parameters are found by fitting a lognormal probability density function, arising from the geometric mean and the  2$\sigma$ (95.5\%) inner and outer bounds.

The second-most compact binary in our sample that displays ellipsoidal variations is the 39.1-min J0106$-$1000. This system is a strong source of gravitational wave radiation; at $i=60.3\arcdeg$, we expect the emission of gravitational waves to cause the orbit to decay at roughly $dP/dt = -1.9 \times 10^{-12}$ s s$^{-1}$ ($-0.06$ ms yr$^{-1}$), which will produce a change in the half-orbital period of $dP/dt = -9.6 \times 10^{-13}$ s s$^{-1}$.

We have constructed an \omc\ diagram of the time-of-minima of the ellipsoidal variations, guided by the period of the highest peak in the FT of our Argos dataset, $39.104063$ min. We find $dP/dt = (0.3\pm6.4) \times 10^{-10}$ s s$^{-1}$, consistent with no change in period, as expected with less than a single year of coverage. Significantly, this effect accumulates with time-squared, so these times of minima will change by more than 10 s within 7 years of our initial observations in 2010~December. Roughly 30~hr of 2m-class-telescope photometry in an observing season yield a phase uncertainty of roughly 8~s, so it is possible to obtain a 3$\sigma$ detection of the spin-up of the tidal bulge due to the emission of gravitational waves within barely a decade of monitoring J0106$-$1000, since the arrival times of the ellipsoidal variations will deviate by $>25$ s after the first 11 years.

We have made a similar set of calculations for the seven other systems with ellipsoidal variations, and include the results in Table~\ref{tab:EVconstraints}. We include the calculated time it would take to make a 3$\sigma$ detection of the period change given the phase uncertainty of 30 hr of 2-m-class photometry, listed as $\tau_{\rm detect}$. It is possible to decrease this detection timescale, since more observations can increase the accuracy with which we can measure the phase of the minima of the ellipsoidal variations. We also include the $T_0$ from the first epoch of observations, which can be used in the future to construct an updated \omc\ diagram with more coverage.


\section{CONCLUSIONS}

We have carried out high-speed photometry of the 20 shortest-period binaries from the ELM Survey \citep{BrownELMv}, all of which contain at least one low-mass WD in a $<4$ hr orbit with another compact companion. Many of these low-mass WDs have high radial-velocity amplitudes, and we detect Doppler beaming in nine of these systems. These signals are generally consistent with the observed radial velocity amplitudes, and we use them to help calibrate and rescale the adopted photometric uncertainties.

More significantly, we detect tidal distortions of eight low-mass WDs in this sample, which we use to constrain the lowest-mass end of the mass-radius relationship for WDs. Unlike typical Earth-sized 0.6 \msun\ CO-core WDs, $<$0.25 \msun\ He-core WDs are similar in size to (and some are even larger than) a giant planet such as Jupiter. There are presently less than 10 other empirical mass-radius determinations for low-mass ($<0.5$ \msun) WDs, and we put our results into context with theoretical mass-radius relations from evolutionary models of He-core WDs.

These models predict that He-core WDs with masses $\leq0.18$ \msun\ should sustain stable hydrogen shell burning (e.g., \citealt{Serenelli02,Panei07,Steinfadt10b}). In fact, a majority of the flux from these $\leq0.18$ \msun\ WDs comes from this residual burning of a thick hydrogen layer \citep{Althaus13}. In addition, unless the systems are perfectly synchronized, tidal heating may also occur, which could effectively heat the primary low-mass WD and inflate it (e.g., \citealt{Fuller12}). Tidal heating may help explain why some of our observed ELM WD radii (such as J0106$-$1000 and J0651+2844) are slightly larger than expected from He-core WD models.

Additionally, a radical change in the structure (and radii) of low-mass WDs between roughly $0.18-0.45$ \msun\ occurs during CNO flashes, which are so-far widely predicted by theoretical He-core WD models \citep{Driebe99,Podsiadlowski02,Panei07,Steinfadt10b,Althaus13}. Such a flashing event may explain the anomalously large radius observed in J0745+1949, which is the coolest tidally distorted WD known but has a radius significantly larger than we would expect given its adopted mass. This WD is bright enough ($g=16.5$ mag) that a suitable parallax distance could confirm such a large radius and better constrain its evolutionary status. If the radius really is 0.176 \rsun, J0745+1949 would be located at a distance of roughly 2.5 kpc. Even at such a large distance, GAIA should contribute a roughly 10-20\% distance estimate \citep{deBruijne12}.

The orbital periods in these systems are shrinking due to the emission of gravitational radiation; all will merge within 6 Gyr, and more than half within 160 Myr. It is possible to use the time-of-minimum of the systems with observed ellipsoidal variations to measure this orbital period decay. The rate of orbital period change depends on the mass of the unseen secondary, which we can estimate from the distribution of $M_2$ from our Monte Carlo analysis of the ellipsoidal variations. Continued observations of these tidally distorted systems enables the exciting prospect of monitoring, on relatively accessible timescales at optical wavelengths, the effects of inspiral of detached, merging binaries as a result of the emission of gravitational wave radiation. Additionally, such observing campaigns afford the opportunity to determine the mass of the unseen companion from a measured rate of orbital decay.

\acknowledgments

We thank the referee M. H. van Kerkwijk for useful comments that greatly improved this manuscript, as well as T. R. Marsh, B. T. G\"ansicke, and E. L. Robinson for helpful discussions. Some of the McDonald Observatory observations were assisted by G. Miller, K. Luecke, A. Rost, J. Pelletier, S. Wang, G. Earle, M. Moore, A. McCarty, and J. Aguilar, undergraduate students in the University of Texas Freshmen Research Initiative. J.J.H., M.H.M. and D.E.W. gratefully acknowledge the support of the NSF under grants AST-0909107 and AST-1312678 and the Norman Hackerman Advanced Research Program under grant 003658-0252-2009. J.J.H. additionally acknowledges funding from the European Research Council under the European Union's Seventh Framework Programme (FP/2007-2013) / ERC Grant Agreement n. 320964 (WDTracer). M.K. acknowledges support from the NSF under grant AST-1312678, and thanks Ben Strickland and Steven Ferguson for useful discussions. M.H.M. additionally acknowledges the support of NASA under grant NNX12AC96G. The authors are grateful to the essential assistance of the McDonald Observatory support staff, especially Dave Doss and John Kuehne. Based on observations obtained at the MMT Observatory, a joint facility of the Smithsonian Institution and the University of Arizona, as well as the Gemini Observatory, which is operated by the Association of Universities for Research in Astronomy, Inc., under a cooperative agreement with the NSF on behalf of the Gemini partnership: the National Science Foundation (United States), the National Research Council (Canada), CONICYT (Chile), the Australian Research Council (Australia), Minist\'{e}rio da Ci\^{e}ncia, Tecnologia e Inova\c{c}\~{a}o (Brazil) and Ministerio de Ciencia, Tecnolog\'{i}a e Innovaci\'{o}n Productiva (Argentina).
	
{\it Facilities:} McDonald 2.1 m (Argos), McDonald 2.7 m (DIAFI), Mt. John 1.0 m (Puoko-nui), MMT 6.5 m (f/5 Science Camera), Gemini-South 8.1 m (GMOS-S)

\section{APPENDIX: NOTES ON SELECTED OBJECTS}

{\em J0106$-$1000}: Our original photometric observations of J0106$-$1000 were published in \citet{KilicJ0106}, announcing what was then the most compact detached WD binary ever known (J0651+2844 was discovered within days of this binary going to press). With just 2.6 hr of Argos photometry on this $g=19.8$ mag WD, we found $1.7\pm0.3$\% relative amplitude ellipsoidal variations. We have followed up those discovery observations with an additional 12.3 hr of photometry using Argos, as well as 4.3 hr using GMOS-S on the 8.1m Gemini-South telescope. These new data confirm the high-amplitude tidal distortions, shown in Figure~\ref{fig:binary1}, and we have measured these variations in three different filters. We refine our original measurement through our typical, broad-bandpass {\em BG40} filter, finding a $1.76\pm0.12$\% amplitude, which we use in Section~\ref{sec:radius} to constrain the WD radius. Using our GMOS-S observations, we find a $1.82\pm0.18$\% ellipsoidal variation amplitude through a SDSS-$g$ filter centered near 4770 \AA\ and a $1.78\pm0.28$\% amplitude through a SDSS-$r$ filter centered near 6231 \AA.

{\em J0923+3028}: We observed J0923+3028, the brightest target in our sample, over three consecutive nights in 2010~December. We detect a modest signal near the orbital period, suggestive of Doppler beaming of the primary. As seen in Figure~\ref{fig:binary2}, the highest peak in the FT does not line up exactly with the RV-determined orbital period. Unfortunately, there is only one brighter comparison star, so we cannot properly explore the impact of atmospheric variability on our observations. Atmospheric variability is likely contributing to the roughly equally significant peak in the FT at 1.4 hr; this signal cannot arise from pulsations of the WD primary, since it is far too hot \citep{HermesELMV45}.

{\em J1741+6526}: This system has the second-highest RV semi-amplitude in our sample, $K_1=508\pm4$ \kms, behind only the 12.75-min J0651+2844. Given the spectroscopically determined mass of the primary, the minimum mass of the unseen companion is 1.11 \msun, and there is a better than 50\% chance that the inclination is such that its companion is more massive than 1.4 \msun. However, this system was not detected in either Chandra or XMM x-ray observations, which likely rules out the possibility of a neutron star companion, requiring the unseen companion to be a massive WD; J1741+6526 is the first confirmed AM CVn progenitor \citep{Kilic14}. We obtained 9.5 hr of photometry in 2011~May and September, which was analyzed in \citet{HermesJ1741}. Our results here include 3.5 hr of additional coverage in 2012~June and July, shown in Figure~\ref{fig:binary2}. The inclination constraints from the ellipsoidal variations, shown in Table~\ref{tab:EVconstraints}, are consistent with a massive WD companion.

{\em J0849+0445}: There are likely some atmospheric effects contributing to inflating the variability at the orbital period (observed with $0.78\pm0.16$\% amplitude), since we expect Doppler beaming to induce a 0.40\% amplitude signal given $K_1$. However, we have only one bright comparison star in the Argos field of view, so we cannot fully constrain the atmospheric contribution to the low-frequency noise, and the uncertainty on our Doppler beaming amplitude is likely underestimated. We do not detect any other significant photometric variability.

{\em J0751$-$0141}: We originally had a difficult time phasing the photometry using the orbital period derived from the RV observations, but an FT of all 63.2 hr of data shows a well-resolved peak at 57.60907 min, which is nearly half the RV-determined orbital period. We thus refined the orbital period to 115.21814 min, which provides for a much more coherent folded light curve, shown in Figure~\ref{fig:binary3}. This is just the fifth known eclipsing ELM WD system. Light curve fits to the shallow primary eclipse find $R_1 = 0.155 \pm 0.020$ \rsun\ and are discussed in \citet{Kilic14}. As with J1741+6526, the inclination constraints from the ellipsoidal variations are consistent with a massive WD companion.

{\em J1234$-$0228}: This binary has the smallest RV semi-amplitude in our sample, with $K_1 = 94.0\pm2.3$ \kms. We obtained more than 8.4 hr of photometry in 2011~January and April. In an FT of all our data, seen in Figure~\ref{fig:binary4}, we see evidence for variability at 76.861 min with $0.30\pm0.06$\% relative amplitude, which is close to but not exactly at the half-orbital period. However, we also see a formally significant alias in the brightest comparison star at 76.824 min with $0.22\pm0.06$\% relative amplitude, so this signal is very likely an artifact from atmospheric variability.

{\em J0745+1949}: The low surface gravity and 8380 K effective temperature of J0745+1949 put it near the instability strip for pulsations in ELM WDs \citep{HermesELMV45}. However, we see no evidence for variability at timescales other than the orbital- and half-orbital periods, to a limit of 0.4\% amplitude. This star also happens to be one of the most heavily polluted WDs known, with deep absorption lines of several different metals that correspond to some of the highest metal abundances observed in any WD. Preliminary analysis of these abundances is presented in \citet{Gianninas13b}.

{\em J0112+1835}: While we expect a 0.34\% amplitude variation at the orbital period corresponding to Doppler beaming, we see no significant evidence for this signal in our 12.8 hr of photometry of this system over four nights in 2011~September. Our three observations are 4.1 hr, 4.0 hr, and 4.8 hr in length, respectively, which makes disentangling a 3.5 hr periodicity more difficult. We have also used this system as a proof of concept to show that RV variations in compact ELM WD binaries are detectable using narrow-band photometry. Motivated by the observational technique of \citet{Robinson87}, we used a custom narrow-band filter with a bandpass centered in the wing of a hydrogen Balmer line to observe this compact binary, with the expectation that RV variations would manifest as periodic variations at the orbital period as the broad Balmer line is shifted into and out of the filter. We took 5.2 hr of observations using Argos through an interference filter centered at 4322 \AA, in the blue wing of the H$\gamma$ absorption line, with a FWHM of 45 \AA. Observing less than two orbits with narrow-band photometry confirms the RV variability ($K_1 = 295.3\pm2.0$ \kms) to the 2.5-$\sigma$ level, as we see a peak at the orbital period in this data at $3.4\pm0.9$\% amplitude. 

{\em J1233+1602}: We obtained 8.8 hr of photometry of this faint ($g=19.8$ mag) system using Argos; our first 2.9 hr run in 2011~May is separated by more than two years from our three runs in 2013~May. Unfortunately we have not covered a complete 3.6-hr orbit, but we have more than 88\% of phase coverage. We see some evidence for a signal corresponding to ellipsoidal variations in this system, shown in Figure~\ref{fig:binary4}, but our detection ($0.61\pm0.22$\% amplitude) is not yet formally significant.


\end{document}